# Development of Quantum InterConnects (QuICs) for Next-Generation Information Technologies

David Awschalom[1], Karl K. Berggren[2], Hannes Bernien[1], Sunil Bhave[3], Lincoln D. Carr[4], Paul Davids[5], Sophia E. Economou[6], Dirk Englund[2], Andrei Faraon[7,8], Marty Fejer[9], Saikat Guha[10,11,12], Martin V. Gustafsson[13], Evelyn Hu[14,15], Liang Jiang[1], Jungsang Kim[16,17], Boris Korzh[18], Prem Kumar[19,20], Paul G. Kwiat[21,22], Marko Lončar[14,15]*, Mikhail D. Lukin[15,23], David A. B. Miller[9], Christopher Monroe[24,25,26], Sae Woo Nam[27], Prineha Narang[14,15], Jason S. Orcutt[28], Michael G. Raymer[29,30]*, Amir H. Safavi-Naeini[9], Maria Spiropulu[31], Kartik Srinivasan[25,32], Shuo Sun[9,33], Jelena Vučković[9], Edo Waks[25,34], Ronald Walsworth[24,34,35,36], Andrew M. Weiner[3,37], Zheshen Zhang[10,38]

[1]Pritzker School of Molecular Engineering, University of Chicago, Chicago, IL 60637
[2]Department of Electrical Engineering and Computer Science, MIT, Cambridge, MA 02139
[3]School of Electrical and Computer Engineering, Purdue University, West Lafayette, IN 47907
[4]Department of Physics, 1500 Illinois St., Colorado School of Mines, Golden, Colorado, 80401
[5]Photonic & Phononic Microsystems, Sandia National Laboratory, Albuquerque, NM 87185
[6]Department of Physics, Virginia Tech, Blacksburg, VA 24061
[7]T .J. Watson Laboratory of Applied Physics, California Institute of Technology, Pasadena, CA 91125
[8]Kavli Nanoscience Institute, California Institute of Technology, Pasadena, CA 91125
[9]E.L. Ginzton Laboratory, Stanford University, Stanford, CA 94305
[10]College of Optical Sciences, The University of Arizona, Tucson, AZ 85721
[11]Department of Electrical and Computer Engineering, The University of Arizona, Tucson, AZ 85721
[12]Department of Applied Mathematics, The University of Arizona, Tucson, AZ 85721
[13]Raytheon BBN Technologies, Cambridge, MA 02138
[14]John A. Paulson School of Engineering and Applied Sciences, Harvard University, Cambridge, MA 02138
[15]Harvard Quantum Initiative (HQI), Harvard University, Cambridge, MA 02138
[16]Department Of Electrical and Computer Engineering, Duke University, Durham, NC 27708
[17]IonQ Inc., College Park, MD 20740
[18]Jet Propulsion Laboratory, California Institute of Technology, Pasadena, CA 91109
[19]Department of Electrical and Computer Engineering, Northwestern University, Evanston, IL 60208
[20]Department of Physics and Astronomy, Northwestern University, Evanston, IL 60208
[21]Department of Physics, University of Illinois at Urbana-Champaign, Urbana, IL 61801
[22]IQUIST, University of Illinois at Urbana-Champaign, Urbana, IL 61801
[23]Department of Physics, Harvard University, Cambridge, MA 02138
[24]Department of Physics, University of Maryland, College Park, MD 20742
[25]Joint Quantum Institute, University of Maryland, College Park, MD 20742
[26]Joint Center for Quantum Information and Computer Science, University of Maryland, College Park, MD 20742
[27]National Institute of Standards and Technology, Boulder, CO 80305
[28]IBM T. J. Watson Research Center, Yorktown Heights, NY 10598
[29]Oregon Center for Optical, Molecular, and Quantum Science, University of Oregon, Eugene, OR 97403
[30]Department of Physics, University of Oregon, Eugene, OR 97403
[31]Division of Physics Mathematics and Astronomy, California Institute of Technology, Pasadena, CA  91125
[32]National Institute of Standards and Technology, Gaithersburg, MD 20899
[33]JILA and Department of Physics, University of Colorado, Boulder, CO 80309
[34]Department of Electrical and Computer Engineering, University of Maryland, College Park, MD 20742
[35]Quantum Technology Center University of Maryland, College Park, MD 20742
[36]Harvard - Smithsonian Center for Astrophysics, Cambridge, MA 02138
[37]Purdue Quantum Science and Engineering Institute, Purdue University, West Lafayette, IN 47907
[38]Department of Materials Science and Engineering,The University of Arizona, Tucson, AZ 85721

*To whom correspondence should be addressed: Marko Lončar: loncar@seas.harvard.edu; Michael G. Raymer: raymer@uoregon.edu

**Abstract**

Just as 'classical' information technology rests on a foundation built of interconnected information-processing systems, quantum information technology (QIT) must do the same. A critical component of such systems is the 'interconnect,' a device or process that allows transfer of information between disparate physical media, for example, semiconductor electronics, individual atoms, light pulses in optical fiber, or microwave fields. While interconnects have been well engineered for decades in the realm of classical information technology, quantum interconnects (QuICs) present special challenges, as they must allow the transfer of fragile *quantum states* between different physical parts or degrees of freedom of the system. The diversity of QIT platforms (superconducting, atomic, solid-state color center, optical, etc.) that will form a 'quantum internet' poses additional challenges. As quantum systems scale to larger size, the quantum interconnect bottleneck is imminent, and is emerging as a grand challenge for QIT. For these reasons, it is the position of the community represented by participants of the NSF workshop on "Quantum Interconnects" that accelerating QuIC research is crucial for sustained development of a national quantum science and technology program. Given the diversity of QIT platforms, materials used, applications, and infrastructure required, a convergent research program including partnership between academia, industry and national laboratories is required.

*This document is a summary from a U.S. National Science Foundation supported workshop held on 31 October - 1 November 2019 in Alexandria, VA. Attendees were charged to identify the scientific and community needs, opportunities, and significant challenges for quantum interconnects over the next 2-5 years.*

# I. Executive Summary

A quantum science and technology revolution is currently in the making, which is widely expected to bring a myriad of scientific and societal benefits. Commensurate with this promise, large challenges exist in seeing the vision become a reality, one of which is the engineering of an essential class of components of any quantum information system—the quantum interconnects.

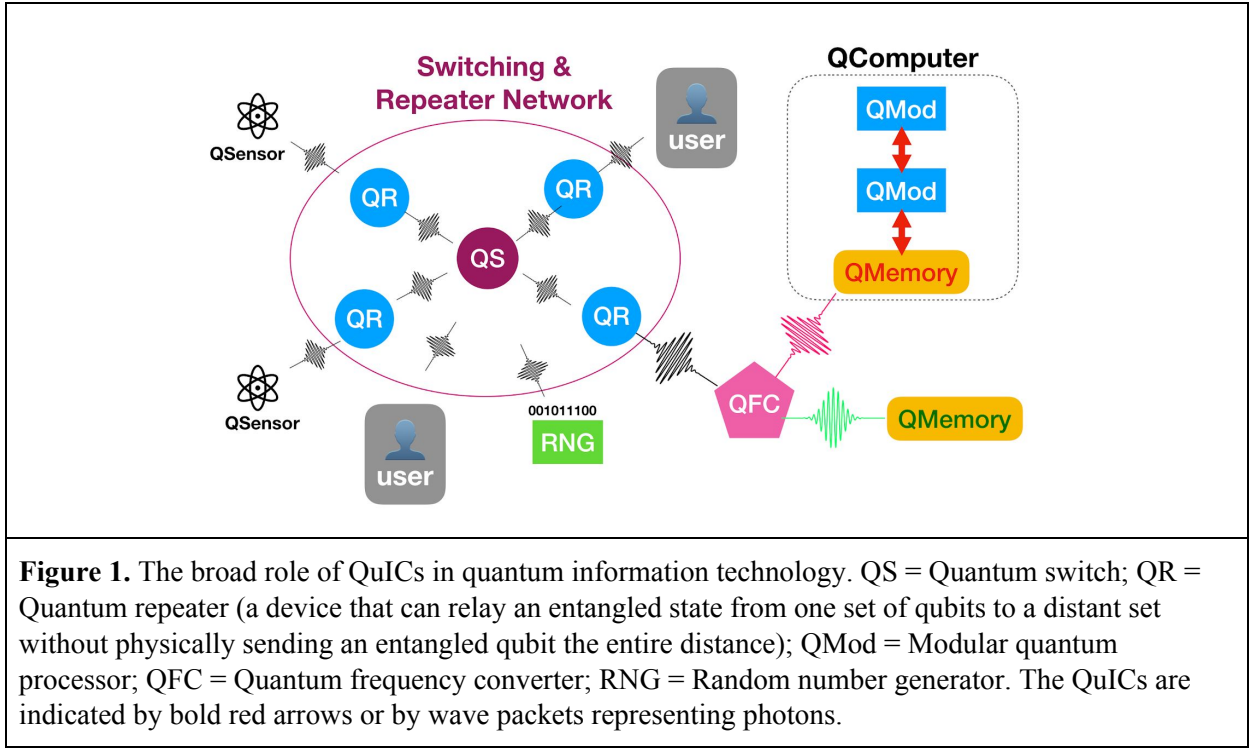

**Figure 1.** The broad role of QuICs in quantum information technology. QS = Quantum switch; QR = Quantum repeater (a device that can relay an entangled state from one set of qubits to a distant set without physically sending an entangled qubit the entire distance); QMod = Modular quantum processor; QFC = Quantum frequency converter; RNG = Random number generator. The QuICs are indicated by bold red arrows or by wave packets representing photons.

Quantum interconnects (QuICs) are devices or processes that allow the transfer of quantum states between two specified physical degrees of freedom (material, electromagnetic, etc.), or, more broadly, connect a quantum system with a classical one. As illustrated in **Figure 1**, QuICs are an integral part of nearly all conceivable quantum information processing systems, including quantum computing, quantum sensing, and quantum communication. For example, it can be argued that modular quantum computing schemes provide the only viable approach that will enable scaling up to truly large numbers of error corrected qubits. Since modular approaches are crucially dependant on efficient QuICs, substantial and focused investment in this vital next stage of quantum computing is timely. Similarly, the ability to transmit information securely by leveraging the laws of quantum physics, in a way that it is "future proof" against even the most powerful quantum computers, is of great national importance. However, the reach of secure fiber-based quantum networks, and the communication rates that they currently allow, are severely limited by the optical losses in the existing quantum interconnects (transmission drops exponentially in conventional optical fibers). Enhancing these interconnects with quantum repeaters will extend the reach and the rate of quantum communication systems. With recent proof-of-principle demonstrations at hand, this effort is ready to be accelerated.

Large technical hurdles exist to implementing QuICs: they must transfer the quantum information (quantum states) with high fidelity, fast rates and low loss, often across a wide range of energies, and do so in a scalable fashion. In some cases a viable candidate QuIC approach is well understood, but dedicated engineering effort is needed to implement it, while in others new physical phenomena need to be explored to implement a given QuIC. An acceleration of research toward the invention and implementation of QuICs will also greatly boost progress in development of materials, devices, systems and supporting infrastructure in critical-path areas that support the development of practical quantum technologies. Such research would enable quantum information science and technology across a wide range of specialties, with ensuing scientific and societal benefits as described in the inset box.

---

**SOCIETAL BENEFITS OF QUANTUM-ENABLED TECHNOLOGY**

In September 2018 the National Science and Technology Council released a report, "National Strategic Overview for Quantum Information Science," which stated, "Through developments in [quantum information science], the United States can improve its industrial base, create jobs, and provide economic and national security benefits."[1] Among the intentions of the national effort outlined by the OSTP report are to: "Focus on a science-first approach that aims to identify and solve Grand Challenges: problems whose solutions enable transformative scientific and industrial progress;" and to "Provide the key infrastructure and support needed to realize the scientific and technological opportunities."

A commentary paper in *Science* [2] co-authored by two participants in the QuIC Workshop, along with a co-author of the NSTC report cited above, summarizes several of the societal benefits that QIT can bring: "A fully functioning quantum computer would radically enhance our capabilities in simulating nuclear and high-energy physics; designing new chemicals, materials, and drugs; breaking common cryptographic codes; and performing more speculative tasks such as modeling, machine learning, pattern recognition, and optimizing hard logistical problems such as controlling the electric energy grid or traffic control systems.[3]" And, "Using qubits instead of conventional bits makes it possible to create shared randomness between parties while knowing whether the communication channel has been compromised by an eavesdropper. This enables sending information securely. Quantum communication can also allow secure communication between multiple parties, and for interconnecting large-scale quantum computers via a quantum internet.[4], [5]" Finally, the *Science* paper also states that the next generation of quantum-based sensors is projected to outperform current sensing technologies, for example in geo-exploration and GPS–free navigation, biological and medical research, and diagnostic technology.

---

**QuIC ACCELERATOR WORKSHOP**
An NSF-sponsored two-day "QuICs Accelerator Workshop" brought together a representative group of over thirty scientists and engineers from academia, industry and national laboratories to identify the present roadblocks that need to be overcome to create functioning QuICs across the necessary range of QIT platforms. The consensus of the participants is that there are concepts and technologies whose development warrants a large, synergistic, and convergent effort involving a range of expertise on a national scale.

## II. Introduction

As quantum technology progresses to real-world applications, a major identified hurdle needs to be overcome: the development of quantum interconnects (QuICs). Just as 'classical' information technology rests on a foundation built of interconnected information-processing systems, quantum information technology (QIT) must do the same. Quantum interconnects include a wide range of systems and processes that allow the transfer of quantum states between two specified physical degrees of freedom (material, electromagnetic, etc.). They may also include components that connect a quantum system with a system that is well described by classical physics for purposes of controlling or reading out information from the quantum system. Quantum interconnects present specific challenges, as they must allow the transfer of fragile *quantum states* between different physical parts or degrees of freedom of the system. With the recent dramatic progress in individual QIT systems for quantum computation, communication, and sensing, an urgent need is to push rapidly toward the integration of such sub-systems to create core technologies that will revolutionize the economy and society in many ways. (See Societal Benefits box).

As quantum systems scale to larger size, a quantum interconnect bottleneck becomes imminent, and surmounting it is emerging as a central goal for QIT. In the context of quantum communication networks, a challenging but extremely important purpose of an interconnect will be to enable the transfer of quantum information (that is, quantum states) across a distance—long or short, depending on the application needs. A prime example of a long-sought-after but elusive subsystem of long-range communication networks (over distances exceeding hundreds of kilometers) is the *quantum repeater*, which would relay an entangled quantum state across a distance that is not accessible using optical fibers only, due to unavoidable signal losses in the communication channels. At shorter length scales, modular quantum computing schemes, which are likely the *only* viable many-qubit near-term approaches, depend crucially on quantum transducers—devices that convert variations in a physical quantity, such as spin state or superconducting flux, into a transmittable signal. Finally, at the chip-scale level, large numbers of quantum memories—devices or systems that can maintain a quantum state over long periods of time—implemented, e.g., using trapped atoms or spin systems in solid state, need to be interfaced using integrated, low-loss and fast on-chip optical networks in order to realize integrated quantum repeaters.

> **QUANTUM ENTANGLEMENT**
> An entangled quantum state describes the joint state (condition) of two or more quantum objects or fields that are statistically correlated in their measured properties, with correlations that are stronger than possible according to classical physics. Entanglement is the essential resource that enables nearly all quantum technology, but is very fragile, making it hard to create and maintain over long times and across large distances.

A consensus in the scientific community is that the technologies needed for quantum computing and quantum networking are closely intertwined, indicating that convergent approaches to these challenges will be the most productive. For these reasons, it is the position of the community represented by participants of the NSF workshop on "Quantum Interconnects" that accelerating

QuIC research is crucial for sustained development of a national quantum science and technology program.

An important affiliated technology is quantum-enhanced sensing of a wide range of physical factors: gravitation, electromagnetism and environmental factors as well as biomedical structure and function. For quantum sensors to reach full capability, in many cases, interfacing them with quantum memories and processors and distributing them across space for collective sensing will be required. Quantum interconnects will play a crucial role in such distributed sensing applications.

> **EXAMPLES of QuICs COMPONENTS:**
> - *communication channel* (optical, acoustic, microwave, etc.) between two quantum systems that can be on the same chip or separated by large distance. Examples include an optical cavity, waveguide or fiber connecting two quantum emitters, or cold microwave waveguide connecting two superconducting-qubit processors;
> - *quantum memory* (e.g., color center, trapped ion, all-photonic cluster state based) and the associated interface to the communication channel;
> - *quantum transducer* used to connect qubits of different kinds (acousto-optical, spin-photon, spin-phonon, etc.), or of the same kind but at different energy (microwave-optical photon, visible-telecom photon);
> - *converter* between different qubit encoding schemes or degrees of freedom (e.g., polarization, temporal, spectral encodings of photons);
> - *small scale & application specific quantum computer,* e.g., quantum repeater, to extend the reach of quantum communication channels;
> - *entanglement sources*—physical processes that create quantum-entangled states of two or more matter-based or photonic qubits.

Combination of different elements of QuIC would enable, for example, links between different processing regions in a quantum computing system in which data qubits are stored in memories (based on, e.g., trapped ions), and transferred into an alternate form (e.g., superconducting qubits) for fast quantum processing. Such hybrid systems will benefit from integrated approaches to connecting classical systems with quantum systems, e.g., for delivery of optical signals to trapped ion- and atom-based quantum computers/clocks/sensors/etc., and to enable efficient data read out.

Finally, it is important to recognize that many information channels, such as an optical fiber or a metallic stripline, largely act as a conduit that can carry both classical signals and quantum states of the signaling medium under appropriate conditions. Thus, classical technologies and quantum-enabled technologies live in a common technological ecosystem with large positive feedback in both directions. For example, classical telecom technology has already provided enormous acceleration of quantum optical communication research; at the same time, the stringent needs of all-optical quantum processors have driven advances in building on-chip reconfigurable multi-mode optical networks, which may benefit classical approaches to information technology. Thus, the *dual-use* paradigm of technology innovation applies to quantum-inspired developments.

## III.A Modular Quantum Processors & Computers

Constructing a large-scale quantum processor is challenging because of the errors and noise that are inherent in real-world quantum systems, as well as the practical engineering challenges that emerge. One promising approach to addressing this challenge is to utilize modularity—a strategy used frequently in nature and engineering to build complex systems robustly. Such an approach manages complexity and uncertainty by assembling relatively small, specialized modules into a larger architecture. Modern high-performance classical computers and data centers are constructed by connecting thousands of computers, memories and storage units into an interconnected network, over which complex computational tasks are distributed. These considerations have motivated the vision of a quantum modular architecture, in which separate quantum systems are incorporated into a quantum network via quantum interconnects [4], [6].

In a modular architecture, the essential building block is the teleportation-based quantum gate, which uses quantum entanglement to connect different modules and thereby implement non-local quantum operations [7]–[10]. In order to connect the modules with each other to perform distributed quantum computation, one has to be able to generate quantum entanglement between pairs of modules to teleport quantum states or quantum gates. Critical figures of merit of such inter-module entanglement generation are (1) the rate of entanglement generation, (2) the fidelity of the generated entanglement, and (3) the reconfiguration of the pairs of modules between which the entanglement is generated.

**Protocols and Progress**

Several protocols have been proposed and demonstrated for transporting quantum information between two nodes. The first is the so-called pitch-and-catch protocol, where a flying qubit, such as a photon, emitted by a stationary qubit (or reflected off a cavity holding a qubit) on the transmitting end would carry the quantum state over the communication channel and transfer it to another qubit on the receiving end [11]. Heroic experiments have been performed using atomic qubits in high finesse optical cavities demonstrating this process [12]. However, the loss in the photonic channel rapidly degrades the performance of this scheme, which makes it impractical at optical frequencies. In superconducting circuits, it is possible to create very strong coupling between the transmitting and receiving qubits with a microwave photon in a transmission line connecting the two modules and featuring negligible loss over the short communication distances involved [13]–[16]. Therefore, such a pitch-and-catch protocol is more practical in these systems.

The second is a heralded entanglement generation protocol, where a pair of entangled qubits in the two modules is first generated probabilistically using photon emission from the qubits and the detection of emitted photons, then a deterministic teleportation of the qubit (or quantum gate) is accomplished using the generated entanglement as a resource. In this protocol, first the communication qubit on each module (such as a trapped ion, neutral atom, atom-like color center in solid-state or quantum dot) emits a photon in such a way that a degree of freedom of the photon (such as polarization, frequency, phase or time-bin, etc.) is entangled with the qubit. The emitted photons are collected (with finite loss), interfere on a 50/50 beamsplitter, and are detected at the outputs. The detection event signals a successful generation of entanglement between the two qubits that emitted the photons. Although the successful execution of the protocol occurs only probabilistically, success is heralded (i.e., confirmed) by detection of two photons at the output of the beamsplitters, and reliable entanglement can be generated at low-to-moderate rates [9], [17].

There have been significant advances in generating entanglement between different modules with improved efficiency and fidelity. In trapped-ion systems, the entanglement generation rate has significantly improved from $10^{-3}$ [18] to ~200 events per second over the course of the past 12 years [19], [20], which enables quantum teleportation between different quantum modules [21], [22]. The advances come from improving the efficiency of photon collection from atoms, reducing photon loss in the channels, and using single-photon detectors with higher detection efficiencies. Similar protocols have been demonstrated in neutral atoms [12], [23], Nitrogen Vacancy (NV) color centers in diamond [24] and quantum dots [25]. In order to ensure that a modular quantum computer can be constructed, it is important to have fully functional quantum computers as the modules, and the entanglement generation rate (quantum communication rate) between the modules must be fast compared to the decoherence rate of the qubits in the modules. Furthermore, efficient optical interconnects to the modules have to be compatible with quantum computing within the modules. For instance, optical cavities can provide an optical interface to atomic quantum computing modules [26] but it remains a challenge to integrate cavities with neutral-atom quantum computing architectures based on Rydberg interactions [27] or trapped-ion quantum computing architectures [28]. Recently efficient quantum optical interfaces have been realized using integrated nanophotonic devices for both trapped neutral atoms[29] and diamond color centers [30].

In superconducting circuits, the pitch-and-catch protocol is indeed practical using a microwave photon as an information carrier. The communication between two superconducting qubit modules has been demonstrated by several research groups [13]–[16]. As long as the communication channel has high quality, it should be possible to send quantum states, even when the number of thermal photons in the channel is much larger than one [31], [32]. Therefore, the current demonstrated approaches can be extended to connecting different dilution fridges using high-quality thermal microwave links.

**Challenges and Research Opportunities**

Recently, proof-of-principle demonstrations of deterministic teleportation-based quantum gates have been carried out in both superconducting-circuit and trapped-ion platforms [33], [34]. These demonstrations show a promising path towards scalable modular quantum computing. However, finding a technical development path to fully modular quantum computers interconnected via quantum communication channels is an extremely challenging task, which requires substantial advances in basic physical principles, device (qubit)-level advances, new protocols, integration of modules and interfaces, and coherent operation across the modules. Here, we outline some of the research directions towards the realization of scalable, modular quantum computers.

**1. Improving quantum interfaces:** While the existing quantum interfaces between modules have seen dramatic improvements, most systems still have not reached the regime where connection between the modules can be utilized for reliable transfer of qubits within the timescale required for distributed quantum computation. For the heralded scheme, we have to continue to improve the entanglement generation rate so that it is comparable to the local entangling gate operation rate within a module. While this is not a strict requirement for efficient quantum computation, it means that the cost of distributing a quantum task across the modules would not substantially constrain the execution of the computational task. Another topic worth noting is that all quantum interconnects are not perfect in terms of the fidelity of the distributed entanglement, or the success probability of the pitch-and-catch scheme. The errors in the quantum interconnects must be minimized or corrected, so that the distributed quantum computation can succeed with viable probabilities, i.e., so that distributing the computation actually improves performance rather than degrading it. New protocols and implementation strategies to overcome the errors in the communication channel need to be developed.

**2. Integration of modules and interfaces:** Seamless integration of the communication interfaces with the computational functions of the modules can introduce some challenges. For example, in heralded entanglement generation protocols, the qubit-photon entanglement generation protocols can lead to decoherence of nearby qubits storing information. For these systems, novel integration approaches must be developed so that the communication and local data processing can co-exist. For solid-state qubits (such as superconducting qubits) that use photons in the microwave range of the electromagnetic spectrum, communication over room-temperature channels becomes impractical. In order to take advantage of modules realized outside the cryogenic environment, frequency up-conversion of the photonic qubit to the optical spectrum is necessary. Quantum transduction techniques to reliably convert microwave photons to optical photons is an important area of research for these applications.

**3. Hybrid modular architectures and interconnects:** The need for modularity can also be driven by the computational functions, where various qubit technologies provide opportunities for executing tasks with different performance requirements. For instance, memory modules that contain qubits with very long coherence times could be implemented on a different platform than processing modules where fast gate times are essential. This potential tremendous advantage comes with additional challenges. In order to take full advantage of such a hybrid modular architecture it is important to develop interconnects capable of distributing entanglement between different qubit implementations, for example, superconducting currents or charges, color centers, neutral atoms, ions, or photons. The spectral characteristics of the photons that couple to each of these systems – including the wavelength and bandwidth – can be very different, by several orders of magnitude, leading to extremely inefficient inter-species conversion in the absence of suitable quantum transducers.

**4.   Coherent operation of modular quantum computer and distributed algorithms**: Even if local quantum computer modules and the needed quantum interconnects are adequately integrated, distributed quantum computation will require operating every module in the system with full quantum coherence among them. This poses challenges in designing and operating phase-coherent control systems across the modules, as well as tracking the quantum phase of every module in the system. Algorithm-level strategies for efficiently distributing the computational task over the modular quantum computer based on the performance specifications of various functional components that constitute the system is therefore an important area of research.

**Table 1 - Timeline and Milestones for Modular Processors**

|  | **3-year** | **5-year** | **10-year** |
|---|---|---|---|
| **Homogeneous Qubit-Qubit Interconnects** | Connection of two fully-functional quantum computer modules with a quantum interconnect. Inter-module entanglement distribution rate better than 10x decoherence rate and 1% of the local gate rate. | Connection of four fully-functional quantum computer modules with a reconfigurable quantum interconnect. Intermodule entanglement distribution rate better than 100x decoherence rate and 10% of the local gate rate. | Manufacturable quantum computer modules with quantum interfaces that can scale to over 100 modules. Intermodule entanglement distribution rate better than 1000x decoherence rate and 100% of the local gate rate. |
| **Transduction with non-native photonic channels** | Demonstration of tunable quantum interconversion between disparate photons (e.g., tunable visible-to-telecom or optical-to-microwave, including bandwidth conversion). | Demonstration of interconversion between microwave and optical photons with high fidelity, SNR, and bandwidth. | Connection to quantum internet. |
| **Heterogenous Qubit-Qubit Interconnects** | Interface between atom- and solid-state-based memory to non-native flexible/tunable photonic channel. | Entanglement between two different types of quantum processor (various atomic and solid-state memory qubits). | QC performance in a multi-node cluster that goes beyond the capability of any individual node, and also beyond those individual nodes connected by a classical network. |

### III.B Quantum Internet

The quantum internet describes a collection of distributed quantum nodes, separated by a range of distances over which one desires to perform some quantum communication protocol that can support, for example, distributed quantum computation (Sect. III.A) or distributed sensing (Sect. III.C). For an accessible overview, see [35]. There are now numerous quantum communication and cryptographic protocols identified, including security distribution for encryption [36]–[43], quantum-certified random number generation in the form of random number beacons and personal devices, secret-sharing [44], [45], quantum fingerprinting [46]–[48] and other multi-party computation protocols, such as secure quantum voting, byzantine agreements, and multi-party private auctions [49]. Of particular relevance is the possibility of "blind" quantum computation [50], [51], whereby a remote user can program a quantum computer without revealing to its owner the algorithm that is run or the computational result, and distributed quantum processing, whereby two or more quantum computers share entanglement to enable them to act as a single larger processor. Because of the distances involved (0.1- 1,000 km), optical photons must be used.

Another key aspect of a fully functioning quantum internet is the potential for unconditional information security—a feature of using quantum information that is not possible with classical information processing. A further benefit of using quantum secured information will be that the lifetime of the security is "infinite"; it will be secure against any advances in computation capability that may occur in the future. There have been many cryptographic tasks in which quantum-secured versions have been conceived. For all of these tasks, quantum interconnects are required because of the need to preserve entangled quantum states.

To realize fully the potential of a quantum internet, significant convergent work is still needed to improve the physical hardware. Theoretical work is also required to develop efficient information processing techniques to preserve the quantum information and determine the most robust and secure network connectivity. The development of quantum-secured devices and protocols could transform the cryptographic landscape.

There are two primary channels over which to transmit the photons: optical fiber and free space. Each of these has challenges and opportunities. The former can leverage the enormous existing network of telecommunication fibers, though then the photons need to be in the telecommunications band to avoid excessive losses. Even still, the transmission through such a fiber will drop exponentially with length, so that direct transmission of quantum states becomes highly inefficient beyond about 100 km. Free-space optical communication is far less well developed, but has the advantage that it can operate over a much larger range of wavelengths, and the losses (due to diffraction) grow only quadratically. Typically, greater care is needed to reduce background light in free-space quantum communication channels; also, there is typically the added challenge of stabilizing the free-space coupling using pointing and tracking methods, and possibly adaptive optics to reduce the effects of turbulence. Nevertheless, many of these challenges have been overcome in a series of free-space quantum communication demonstrations, between mountains [52], [53], over water [54] within cities [55]–[57], from airplanes [58], balloons, and drones [59] and even using satellites in low-earth orbit [60], [61]. While the achieved transmission rates in these experiments might have greatly exceeded what would have been possible using fiber channels—in one case by nearly 20 orders of magnitude [60], [61], they are still often very low, and methods such as multiplexing or employing higher-dimensional states (see below) may be needed to achieve practical rates.

**Challenges and Research Opportunities**

To build a fiber-based global network capable of distributing quantum entanglement, there are two main challenges that have to be overcome. First, optical attenuation during fiber transmission leads to an exponential decrease in the entangled-pair distribution *rate*. Second, operational errors such as channel errors, gate errors, measurement errors, and qubit memory errors can severely degrade the *quality* of the distributed entanglement, which at best reduces the quantum advantage and at worst completely eliminates it, e.g., a quantum cryptographic key may be completely insecure!

**1. Quantum repeaters:** To overcome these challenges and extend the range of fiber-based entanglement distribution beyond a few hundred kilometers, quantum repeaters (QRs) are required, but are not yet available. Depending on the tools used for suppressing these imperfections, the quantum information community has identified the following three generations of QRs: The first generation of QRs [62], [63] uses heralded entanglement generation and heralded entanglement purification, which can tolerate more errors but requires two-way classical signaling over the entire chain of QRs; such signaling then implies that the requisite quantum memory lifetimes/coherence times must be substantially longer than the round-trip communication times. The second generation of QRs introduces quantum encoding and classical error correction to replace the entanglement purification with classical error correction, handling all operational errors [64], [65], which is more demanding in physical resources but requires only two-way classical signaling between neighboring repeater stations, and consequently further improves the quantum communication rate. The third generation of QRs would use quantum encoding to deterministically correct both photon losses and operation errors [66], [67]. By entirely eliminating two-way classical signaling, the third generation of QRs would promise extremely high entanglement distribution rates that can be close to classical communication rates, limited only by the speed of local operations, in turn limited by, e.g., photon source rates, detector saturation rates and timing jitter, etc.

One important benchmark for QRs is the repeater-less bound [68], [69], which imposes the fundamental limit of the direct quantum communication protocols. Recently, there have been significant advances in experimentally demonstrating key elements of a QR in an integrated system. An important recent highlight is the experimental demonstration of memory-enhanced quantum communication surpassing repeterless-less bound in proof-of-concept laboratory setting, using a solid-state spin memory associated with Silicon Vacancy (SiV) color center integrated in a diamond nanophotonic resonator [30], [70]. This paves the way towards the demonstration of a full quantum repeater, which in turn will enable scalable large-scale quantum networks.

**2. Quantum memories:** The major challenge for the first generation of quantum repeaters is the development of long-lived quantum memories with efficient optical interfaces, such as addressable color center nuclear spins with integrated nanophotonics [71], trapped-atomic qubits with Purcell-enhanced emission [12], [19], [23], or superconducting circuits with microwave-to-optical transduction. In addition, the availability of efficient photon detectors with low dark counts is crucial, with significant advances needed in reducing the cost, integration, etc.

**3. Spectral-temporal encoding:** It is now generally recognized that practical rates of entanglement distribution can likely be achieved only by employing high levels of channel multiplexing (e.g., spectral, temporal, spatial) to enhance success probabilities; for instance, quantum signals are simultaneously sent at multiple nearby wavelengths or in multiple time bins. Although each spectral or temporal channel has some probability of failure or loss, the likelihood that all would be unsuccessful decreases with the number of multiplexed channels. However, one needs a mechanism to demultiplex into a single spectral-temporal mode; alternatively, they are each coupled to their own quantum memory qubit, but then some mechanism for identifying and coupling a particular pair of successfully "loaded" quantum registers is needed. The use of such temporal multiplexing has recently enabled a 30x enhancement in the success rate of a two-photon quantum communication protocol [72]; the advantages become exponentially larger for protocols requiring higher numbers of qubits. The benefits of multiplexing arise only if the quantum interconnects that implement the multiplexing and demultiplexing have high fidelity and low loss.

Another emerging strategy is to use qudits, the higher-dimensional counterparts to qubits, e.g., using three time bins to encode numbers 0, 1, and 2, and arbitrary superpositions thereof. Just as it does for classical communication, such encoding increases the information-carrying capacity of a photon by $\log(d)$ where $d$ is the dimensionality, at the expense of more complex measurements and manipulations. Finally, encoding multiple qubits (or even their higher-dimensional counterparts, qudits) onto a single photon can yield intrinsic robustness to loss: because all of them are guaranteed to be lost or transmitted together, the net success probability can be greatly enhanced. For instance, the probability that a channel with 99% loss will successfully transmit a 3-photon three-qubit state is only $10^{-6}$; in comparison, a single-photon three-qubit state experiences the loss only once, i.e., with a 1% success probability. The concept of qubit entanglement also generalizes to hybrid entanglement [73], between different degrees of freedom of a single photon, e.g., polarization and spatial mode, and hyper-entanglement [74], between multiple corresponding degrees of freedom of two photons, e.g., polarization and time bin [71], or time-bin and frequency-bin [75], [76]. One critical need is a method to transduce such higher dimensional quantum states into qubit memories.

**4. Efficient measurements:** Finally, all three generations of QRs can be greatly enhanced by including efficient quantum non-demolition (QND) measurements [77] – a measurement that records the successful passing of a photon without observing it or changing its quantum state. In this way, any memory can be converted to a *heralded* quantum memory, which enables one to know whether a photon has successfully been transmitted down the entire length of a communication channel; such knowledge greatly reduces the required number of quantum memories, since one is only needed in cases where the quantum signal was successfully transmitted through the optical channel.

With the emerging demonstrations of quantum repeaters, it will be important to optimize them to overcome realistic imperfections through use of robust architecture and encoding. It is also urgently needed to develop novel quantum network applications and appropriate corresponding performance metrics, such as entanglement fidelity, throughput, latency, resource overhead, etc. These performance metrics should also guide the device design and fundamental investigation of relevant physical platforms.

**Table 2 - Timeline and Milestones for Quantum Internet**

|  | **3-year** | **5-year** | **10-year** |
| --- | --- | --- | --- |
| **Major Achievements** | Detected photonic entanglement rate beyond $10^8$ ebits/second | Quantum repeaters with error correction against operation errors | Forward error-corrected photonic quantum states for one-way repeaters |
| **Distance and Rates** | Entangled quantum memory over > 10 km distance | Verifiable quantum entanglement distribution over >100 km at > 1 M-ebits/sec; distillable entanglement rates >100k-ebits/sec | Quantum networks reaching transcontinental scales of thousands of km |
| **Capability of Repeater Nodes** | Quantum repeater node via entanglement swapping beyond direct transmission | Active error correction against operation errors; many-party protocols demonstrated in fielded quantum networks | Full error correction against loss and operation errors; hybrid nodes with different functions. |
| **Number of Repeater Nodes** | Quantum networks with >3 memory nodes and >10 user nodes | Networks of >10 quantum repeaters/quantum computers in superposition | Networks with >100 of repeater nodes |
| **Free-Space Quantum Network** | Constellation of 3-5 mobile platforms demonstrated | Entanglement swapping between space-earth | Transcontinental entanglement distribution via quantum memory-enabled satellite |
| **Quantum Network Applications** | Quantum-secured communication rate exceeding 1 MB/sec over 100 km | Network-based quantum metrology | Blind Quantum Computing |

## III.C Quantum-Enhanced Sensors

Quantum-sensing technology has made significant progress over the last few decades and has given rise to atomic clocks [78], magnetometers [79], and inertial sensors [80] that operate at the standard quantum limit (SQL). With the tremendous advances in the theoretical and experimental aspects of quantum information science over the last decade, new quantum resources, such as quantum memories and entangled particles, can now be harnessed to enhance further the performance of quantum sensors. Also known as quantum metrology, quantum-enhanced sensing is aimed at taking advantage of these emerging quantum resources to outperform the SQL and achieve unprecedented sensing performance. As a remarkable instance of quantum-enhanced sensing, the Laser Interferometer Gravitational-wave Observatory (LIGO) utilizes non-classical squeezed light to enable a measurement sensitivity below the SQL [81]. Quantum-enhanced sensing has also been proven to be a powerful paradigm for a variety of scenarios including magnetic sensing with quantum memories [82], quantum-illumination target detection [83], sub-SQL atomic clocks [84], and nano-mechanical sensors [85].

Most existing quantum-enhanced sensing demonstrations leverage non-classical resources to improve the measurement performance at a single sensor, but many real-world applications rest upon a network of sensors that work collectively to undertake measurement tasks. Notable examples for such a setting include wireless sensor networks [86], phased arrays [87], and long-baseline telescopes [88]. In this regard, the quantum internet presents unique opportunities for quantum sensors to utilize shared entanglement to boost the performance in networked sensing tasks. The following section discusses the concept, promising research avenues, and application space for interconnected quantum sensors.

**Interconnected Quantum Sensors**

Extensive studies have been dedicated to using bipartite (two-party) entanglement as a resource to overcome the SQL at a single sensor. In one step forward, recent theoretical works on quantum-enhanced sensing based on multipartite entanglement show that interconnecting distributed quantum sensors to form an entangled sensor network can probe global parameters at the Heisenberg limit, i.e., at an estimate uncertainty that scales favorably compared to the scaling for a network of independent sensors. Specifically, Ref. [89] proposed a quantum network of clocks that enjoys boosted precision and security over conventional classical clock networks. More generally, two theoretical frameworks for distributed quantum sensing based on, respectively, discrete-variable [90], [91] and continuous-variable multipartite entanglement have been formulated [92]. On the experimental front, a proof-of-concept distributed quantum sensing experiment demonstrated the utility of multipartite continuous-variable entanglement for enhancing the measurement sensitivity for estimating global phase shifts. To demonstrate the prospect for interconnected quantum sensors in real-world applications, Ref. [93] reported an entangled radiofrequency (RF)-photonic sensor network in which distributed RF sensors harness their shared multipartite entanglement to enhance the precision of estimating the properties, e.g., the angle of arrival, of an incident RF wave across all sensor nodes.

In the context of a quantum internet, quantum sensors distributed over a distance will be able to establish high-fidelity entanglement to achieve measurement sensitivities beyond the SQL.

Potential application scenarios for large-scale entangled quantum sensor networks would encompass high-precision astronomical observation [94] [88] environmental and health monitoring, positioning, navigation, and timing. Two possible means of building up entanglement shared by quantum sensors are the following: 1) a matter-based quantum sensor first entangles with a photonic mode, which is then transmitted through the quantum internet equipped with quantum repeaters to ensure high-rate long-distance entanglement distribution. Entangling photonic quantum measurements are performed at the destination quantum repeater nodes to establish multipartite entanglement between matter-based quantum sensors. 2) As an alternative method to form an entangled quantum sensor network, photonic multipartite entanglement tailored for a specific networked sensing task is first produced by a photonic quantum chip at a central node. Each arm of the photonic entangled state is then transmitted to a quantum sensor located in the quantum internet. As in the matter-based quantum sensor network, the quantum internet takes advantage of quantum repeaters to compensate for entanglement distribution loss so that high-fidelity photonic multipartite entanglement is maintained. At each quantum sensor node, a high-efficiency low-noise quantum transducer converts the information carried by the object of interest into the photonic domain so that quantum measurements on the photonic multipartite entanglement unveil the global property of the interrogated object.

**Challenges and Research Opportunities**

Apart from the need for a quantum internet as a backbone, a number of technical accelerations will be critical for the construction of entangled quantum sensor networks.

**1. Device concepts:** Matter-based quantum sensor networks are comprised of any of a diverse range of useful sensors. Examples, not exhaustive, include sensors of massive particles, photons, magnetic fields, electric fields, temperature, gravity, pressure, and chemical processes. Such sensor networks require efficient light-matter interfaces or interconnects to create entanglement between quantum sensors and photonic modes. In an ideal situation, establishing entanglement between multiple matter-based quantum sensors at the quantum repeaters calls for deterministic multipartite Bell measurements with near-unity efficiency. Such a measurement can be realized by first transferring the quantum states of photons into those of solid-state qubits, followed by fault-tolerant quantum computation on a special-purpose small-scale quantum computer. As a necessary ingredient, the outcomes of the Bell measurements need to be communicated to different quantum-sensor nodes in realtime, by fast electronic processing and a low-latency classical communication network. Since most matter-based quantum sensors operate with readout in the visible to the near-infrared spectral range, high-efficiency low-loss quantum frequency converters are required to shift the wavelengths of photons into the telecommunication window for long-haul communication via a quantum internet.

The device requirement for the photonic quantum sensor network encompasses envisaged programmable photonic quantum chips (PQCs) to generate appropriate photonic multipartite entangled states. Each PQC would entail low-loss waveguides and couplers, high Q-factor ring resonators, and single quantum emitters that provide needed ('non-Gaussian') resources for universal quantum information processing. The produced photonic multipartite entangled states need to be inserted into optical fibers through couplers with near-unity transmissivity. The PQCs should also provide classical controls to pre-compensate the dispersion and other imperfections incurred in the transmission. At each quantum sensor node, high-efficiency quantum transducers convert the physical information contained in the microwave, mechanical, or magnetic domains into modulations on the visible photonic quantum states. To ensure high performance for the quantum sensor network, it is important for the quantum transducers to achieve high efficiency while minimizing additional loss and noise.

To extract information carried by the photons, high-efficiency quantum-limited homodyne, heterodyne, or direct measurement detectors are subsequently utilized. These detectors ideally will be integrated on the same PQC as are the quantum transducers to obviate additional coupling and conversion losses. Recent advances in the fabrication and integration of quantum devices based on widely used optoelectronic materials such as lithium niobate [95] point to a promising platform for interconnected quantum sensors. Further technology accelerations would lead to a versatile photonic quantum sensing platform capable of accommodating hundreds to thousands of elements with different functionalities on the same PQC.

*2.* **Sensor network architectures:** *architectural perspective*, the engineering of multipartite entangled states for a large-scale quantum sensor network comprised of a large number of sensors remains an open problem, due to the complexity of multipartite entanglement. In this regard, machine-learning tools would be useful for identifying near-optimum entangled states for networked sensing problems. A recent theoretical study shows that the optimum entangled state and measurement configuration can be found by training photonic quantum circuits by a support-vector machine and a principal component analyzer, for data classification and data compression tasks at a physical layer [96]. Further investigations would incorporate the machine-learning framework into quantum devices and the quantum internet under development to accelerate the performance, scale, and application scope.

**Table 3 - Timeline and Milestones for Quantum Sensors**

|  | **3-year** | **5-year** | **10-year** |
|---|---|---|---|
| **Matter-based quantum sensors** | Entanglement-enhanced sensing in local registers (e.g., multi-nuclear or electron-nuclear spin entanglement around or within a color center). | Entanglement-based distributed solid-state sensors entangled leveraging spin-photon entanglement. Demonstration of entanglement assisted clock synchronization (in one room - easier, across world - geography is a problem). | Sensors connected to quantum internet. |
| **Photonic quantum sensors** | Scale up to 10 entangled photonic sensors in an integrated platform. Development of various transducers including high-efficiency RF-photonic and optomechanical transducers. Improvement of chip-to-fiber coupling efficiency to ~80%. Applications include RF, inertial, mechanical sensing, etc | Fully reconfigurable on-chip entangled photon sources operating at different frequencies and entangled degrees of freedom. Integrated on-chip transducers > 95% chip-to-fiber coupling efficiency. New entanglement-enhanced sensing approaches for classical noise rejection, high resolution, etc. Entanglement-enhanced multi-aperture telescopy | Scale up to ~100 entangled sensors. Integrate with the quantum internet for long distance entanglement distribution to sensors. Use quantum error correction to enhance sensitivity. Incorporating entangled sensors into existing classical sensing infrastructures. Long-baseline telescope enabled by quantum repeaters. |

## IV. Convergent Acceleration Opportunities

An acceleration of technical research toward invention and implementation of quantum interconnects will greatly boost progress in quantum information science and technology across a wide range of specialties. Particularly important needs and goals are summarized here.

### IV.A Devices and Systems

Future quantum computers and networks will require unprecedented connectivity over distances ranging from micrometers to hundreds or even thousands of kilometers. Such diverse connectivity will put significant demand on quantum interconnects, requiring, for example, the scalable fabrication and integration of a large number of components in compact opto-electronic chips.

In particular, integrated (on-chip) device technologies are likely to play a number of important roles in implementing quantum interconnects. The general ability to enhance interactions through control of the electromagnetic density of states in suitably engineered geometries enables a wide variety of physical resources to be realized, while the manufacturing technologies used to create such geometries can be predicted to reach the level of scaling and integration required. Here we outline different QuIC technologies needed, and then discuss specific device and material platforms in which these technologies should be developed.

Quantum interconnects will provide the links between various quantum devices to realize large-scale systems. We organize interconnects into three primary categories:

1) *Interconnects between bosonic and atomic systems*: The role of such interconnects is to interface bosonic fields with atoms for a variety of applications. Here "bosonic fields" include a variety of harmonic-oscillator systems that are suitable to carry quantum information across distance, such as optical photons, mm waves, microwaves, and acoustic phonons. "Atomic" systems should be broadly interpreted to encompass all forms of matter systems including neutral cold atoms, trapped ions, Rydberg-excited atoms, cold molecules, quantum dots, color centers, impurity bound excitons, superconducting josephson devices, etc. Examples include transfer of quantum information from atomic memories to photons for quantum networks, readout of phase information in quantum sensors, transfer of microwave excitations from transmon qubits to microwave cavities, etc.
2) *Interconnects between two bosonic photonic platforms*: The role of these interconnects is to impedance-match two bosonic platforms or information-encoding schemes in order to achieve interconversion or to combine various quantum systems. Examples include quantum frequency conversion of optical and microwave signal to the telecom, temporal-waveform conversion that enables optimal interfacing of heterogeneous quantum nodes, quantum transduction between optical photons and acoustic phonons, and connections between photonic systems that use different encodings (e.g., time-bin, frequency-bin, wave-packet shape, coherent or squeezed-state, or polarization-state).
3) *Interconnects between two atomic platforms*: Here, quantum interconnects mediate long-range interactions between atomic systems. Examples include entanglement of quantum memories separated by large distances, hybrid quantum systems composed of different matter qubits [97], microwave interconnection of transmon qubits in superconducting devices, development of low-loss switches and architectures to connect arrays of quantum systems to each other in a scalable fashion, etc.

Below we describe the various requirements for these different QUIC applications.

**1. Atomic-photonic interconnects**

Quantum information relies on a broad array of matter-based quantum systems to store and manipulate quantum coherence. Such matter systems include single atoms [27], quantum dots [98], color centers [24], [99], [100], rare-earth ions [101]–[106], defect-bound excitons, and superconducting josephson devices [107], to list a few. These systems provide a variety of essential functionalities in quantum information that include single-photon sources [108], quantum memories [101], [109], quantum sensors, and photon storage devices.

An essential role of quantum interconnects is to interface these matter systems with optical photons, the unique carriers of quantum signals across long distances. Interconnects used to form shorter-distance networks, e.g., inside a room, or on a chip, may use lower-frequency photons (e.g. mm-wave [110] or microwave [111]) or even acoustic phonons [112] to accomplish the analogous connectivity function. The quantum interconnect must provide strong interactions in order to mediate quantum state transfer, entanglement, or other uniquely quantum resources. Typically, such functionality entails using optical cavities or waveguides to enhance light-matter interactions into the single-photon regime. These photonic structures must support high quality factors and small mode volumes to attain the desired interaction strengths. Furthermore, these devices often have to operate at short wavelengths (visible and near-IR) which puts additional constraints on the materials used. Another important requirement of such interconnects is low insertion loss. Low-loss operation is particularly important in quantum applications where the loss of a single photon can destroy the quantum state of the entire system. Finally, there is the issue of compatibility of the interconnect hardware with the atomic systems. Many atomic systems operate in millikelvin cryogenic environments with miniscule acceptable heat loads, and extreme sensitivity to quasi-particles generated by optical absorption. Scalable atomic-photonic interconnects must be able to scale within these constraints.

**2. Photonic-photonic interconnects**

The future quantum internet will likely be composed of a broad range of disparate systems that must interact and exchange quantum signals. The most viable candidate for interacting different quantum systems across a distance is via photonic channels. But these diverse quantum systems emit photons with diverse frequencies and temporal shapes and durations, necessitating quantum interconnects that couple photons with different properties. Important examples include quantum frequency conversion to telecom, microwave-to-optical conversion, and spectral bandwidth conversion.

The choice of interconnects largely depends on the distance scale over which such quantum interconnects operate. Microwave photons may be suitable for intra-chip and inter-chip links within a mK environment where thermal background is suppressed, but are unlikely to be used for connections over much longer length scales. Telecommunication-band photons remain the information carrier of choice for long-distance networks based on optical fiber, while long-distance free-space links and shorter metro-area networks may be amenable to optical photons in different frequency bands. While most visible photon frequencies will be suitable for relatively short distance networks (e.g., to connect nodes within a distributed quantum computer), there is also the potential to work with millimeter-wave and terahertz frequency photons for sufficiently short links.

Quantum frequency conversion (QFC) devices, e.g., based on three- or four-wave nonlinear optical mixing [113], [114], [115], [116] or direct frequency shifting using electro-optic modulation [117], enable the spectral translation of a quantum state of light to targeted frequencies with high efficiency, low added noise, and sufficient bandwidth. QFC devices are needed to enable quantum interconnects that link the most suitable matter qubits for a given application across the relevant length scales and photonic communication medium [118]. For networks consisting of homogeneous nodes, these QFC devices may primarily consist of downconversion and upconversion units that are ideally seamlessly integrated with the photonic qubits that are directly coupled to the matter qubits. For networks consisting of heterogeneous nodes, it is likely that QFC needs to be combined with coherent temporal-spectral waveform manipulation, to ensure optimal coupling to matter qubits that may have significantly different acceptance bandwidths and lineshapes.

Another approach to linking matter qubits over distance is through the use of intermediate entangled-photon-pair sources that are engineered to create one photon at a frequency suitable for direct interaction with the matter qubit (e.g., at 637 nm for an $NV^-$ center in diamond), and another at the relevant frequency for transmission across the physical interconnect channel (e.g., 1550 nm for a long-distance fiber link). Through entanglement swapping, such sources can be used to entangle distant quantum nodes, though if those nodes are heterogeneous, it is likely that some form of waveform reshaping will also be needed, e.g., using "time-lensing" methods [119].

Multiplexing and demultiplexing of optical pulses based on their temporal mode identity (wave-packet shape) can also play an important role [120]. As mentioned above, by providing a high-dimensional state space for single-photon packets, temporal modes offer higher information content per photon. The ability to demultiplex light into temporal-mode components offers increased signal-to-noise ratio in photon-starved communication links, such as can be envisioned in deep-space communication [121].

### 3. Atomic-atomic interconnects

Given that near-future quantum computers and networks will likely be composed of modular units of a few tens to hundreds of matter qubits (atoms, solid-state defects, superconducting devices, photonic, etc.) that can perform small-scale quantum information tasks, scaling to larger systems will require interconnection of multiple modular components. Photonics provides an ideal approach to achieve this interconnectivity. For example, current state of the art in fully controlling trapped-ion qubits involves of the order 30 ions; scaling up to hundreds of ions will require using light for interconnecting separate modules each containing around this number of ions. Industrial quantum computers have recently reached 53 qubits [122], but scaling up to hundreds to thousands of qubits on a single chip is presently well out of reach.

A scalable optical interconnect for atomic modular nodes requires the ability to route photons efficiently and provide high-fidelity multi-photon interference. This multi-photon interference provides the necessary quantum step to generate effective atom-atom interactions at a distance.

Interconnects should ideally combine photonics with detectors and other components to provide a complete on-chip solution for large- scale modular quantum information processing. Additional functionality such as filtering, on-chip quantum frequency conversion, and multiplexing will significantly enhance the scalability of the total system. In many other circumstances where the atoms possess large material-strain susceptibility [123], acoustic phonons might be particularly suitable to mediate interactions between two or multiple atomic qubits on a chip [124].

## 4. Integrated quantum photonics platforms

*Atomic-photonic interconnects*

Several different photonic platforms naturally host atomic-like systems, including diamond (color centers) [82], GaAs and InP (quantum dots), and SiC (color centers) [108], [125] and various transparent crystals doped with rare-earth ions [101], [104] or transition metals. Integrated devices are important for realizing the crucial local interconnect between a matter qubit and a photonic qubit. For solid-state qubits, this often involves engineering of suitable photonic cavities or waveguides to ensure that, for example, emitted photons entangled with spins are efficiently funneled into a single desired collection channel. It is critically important that the fabrication processes that create such photonic structures, which sometimes have features at the 100-nm-length scale, do not induce excess dephasing or spectral diffusion that will limit the coherence properties of the matter qubit. Such issues are increasingly being addressed through design constraints on the separation of the matter qubit from an etched surface, surface passivation techniques, and electrical charge stabilization methods. Hybrid integration of materials that host quantum emitters with materials that support scalable fabrication of photonic devices [126] may also be beneficial to the atomic-photonic interconnects.

For photonic chip integration of trapped neutral atoms, ions, and cold molecules, protecting qubit coherence in a platform with a wide enough transparency window to accommodate the short wavelength photons associated with these systems is paramount. For trapped ions, this may require new electromagnetic designs that limit deleterious effects on the electrostatic traps; similar approaches may be needed for systems like Rydberg atoms, which are unlikely to be brought close (within the evanescent tail) to a guided-wave photonic structure. Other atomic systems, in particular single neutral atoms and cold molecules, can be loaded and trapped within the evanescent field of photonic devices, enabling an efficient matter-photon qubit interconnect. Recent progress with optical tweezer traps in Rydberg chains and small collections of neutral atoms and molecules offer a key challenge for quantum interconnects.

*Photonic-photonic interconnects*

Several different nonlinear nanophotonic platforms are being developed within the QuIC community to enable the QFC and entangled photon-pair interconnect approaches needed to realize photonic interconnects across different frequency bands. In practice, the most mature technology is based on cm-scale quasi-phase-matched nonlinear media [127] such as periodically-poled lithium niobate (PPLN) waveguides, where internal conversion efficiencies approach 100% and signal-to-background levels in excess of 100:1 are achievable for single-photon-level inputs. In its current state, this waveguide technology is not directly amenable to dense integration within photonic circuits (the optical mode field diameter and bend radius are similar to those of an optical fiber), nor is it directly amenable to direct integration with single quantum emitters or a variety of other integrated photonics technologies. Its further development is important from the perspective of providing hardware for near-term networking efforts and for implementing strategies that realize waveform conversion via nonlinear optics.

Integrated nanophotonic platforms based upon geometries with high refractive index contrast (typically created using thin-film nonlinear optical materials on a low refractive index substrate such as $SiO_2$) can enable QFC and waveform conversion approaches using manufacturing approaches that can be highly scalable and enable complex integration with other photonic circuit functionalities, including beam splitters and filters . For second-order nonlinear processes, thin-film lithium-niobate-on-insulator (LNOI) [128], [129] and AlN [130] have shown particular promise, given their wide optical transparency window, very low optical losses [131], [132], appreciable nonlinear coefficients, and amenability to nanofabrication processes. For third-order nonlinear processes, silicon nitride has risen to the forefront of many related classical applications (e.g., compact frequency comb generation), and QFC of quantum dot single-photons and photon-pairs using silicon nitride nonlinear resonators has recently been demonstrated [116], [133]. III-V semiconductors such as GaP [134] and AlGaAs and wide-bandgap materials such as SiC [100] and diamond also possess strong optical nonlinearities [135], but have not yet achieved the level of performance of the aforementioned systems. However, their ability to directly host matter qubits (e.g., color center or quantum dot spins) is of significant benefit to integration.

For all of these platforms, the basic strategies for achieving efficient frequency conversion are generally conceptually well-understood, and fabrication techniques are developed. However, there are still many challenges in realizing connections between the ultra-wide frequency separations [136] (e.g., UV-telecom), and understanding the relevant noise generation mechanisms (impurity-based fluorescence, Raman scattering, spontaneous parametric processes, to name a few) is an ongoing process [137], [138].

Materials supporting a second-order nonlinearity often exhibit an appreciable electro-optic effect [95], which enables fast (tens of picoseconds) reconfigurable switching operations. For scenarios in which slower speeds are adequate (e.g., MHz switching bandwidths), thermo-optic and micro-electromechanical switches can be considered, though the former do not function well at cryogenic temperatures.

Microwave-to-optical QFC typically requires access to physical processes distinct from those described above, with the exception of electro-optic platforms [139] that do provide natural links between the two frequency bands, though the extent to which such links can be sufficiently low noise in practice is not known. Piezoelectric media [112], [140] such as LNOI, AlN, GaP, and GaAs are being considered for microwave-to-optical QFC mediated by nanomechanics, while modular approaches based on free-space cavities [141] coupled to electromechanical systems have shown the best performance thus far in terms of efficiency, albeit over moderate bandwidths and without adequately low noise. Another way to mediate the interaction between optical and microwave photons is via atomic ensembles simultaneously coupled with high cooperativity to photonic and microwave resonators. In this context, rare-earth-doped media are particularly well suited as microwaves can be coupled to either Zeeman or hyperfine transitions in ensembles exhibiting very narrow inhomogeneous lines[142], [143]. Unlike QFC between optical wavelengths, there has to this point been no full demonstration of QFC between microwave and optical wavelengths where, for example, non-classical photon statistics or quantum interference are shown to be preserved.

From the above, it is evident that a wide variety of material platforms are likely needed to address the full range of quantum interconnect challenges. One approach to combining the best attributes of multiple systems is heterogeneous integration of multiple materials into a common platform. For example, rather than developing new QFC resources in III-V materials or diamond, heterogeneous integration [144] with $Si_3N_4$, AlN, or LN would enable a direct coupling of the matter-photonic qubit interface with the QFC interface. Several approaches, including full wafer bonding, die bonding, transfer printing, and pick-and-place device transfer, are being considered by the community to realize this functionality.

**Table 4. Integrated Quantum Photonics Platforms**

| Platform | Transparency window | Nonlinear coefficient | Demonstrated optical loss | Single quantum emitter integration | Qubit integration | Tuning/ Switching mechanism |
|---|---|---|---|---|---|---|
| **Silica** | >140 nm | weak chi(3) | Ultra-low | Not native | Not native | Thermo-optic |
| **Silicon nitride** | >350 nm | Moderate chi(3) | Low | Not native | Not native | Thermo-optic/Micro-electro-mechanical-systems (MEMS) |
| **Silicon-on-insulator** | >1000 nm | Strong chi(3) | Medium | Nascent (Se defects) | Electron spin qubits | Thermo-optic/ free-carrier/ MEMS |
| **LiNbO3-on-insulator** | >300 nm | Strong chi(2); moderate chi(3) | Low | Rare earth incorporation | Not native | Electro-optic/ piezo-electric |
| **AlN** | >200 nm | Moderate chi(2); moderate chi(3) | Low | Not native | Not native | Electro-optic/ piezo-electric |
| **GaAs-on-insulator** | >750 nm | Strong chi(2); strong chi(3) | Medium | InAs quantum dots | Single electron/hole spins | Piezo-electric |
| **SiC-on-insulator** | >400 nm (4H) | Moderate chi(2); moderate chi(3) | High | Via electron beam irradiation, ion implantation | Single electron/ nuclear spins | Electro-optic, piezo-electric, DC Stark-shift |
| **Diamond-on-insulator** | >250 nm | Moderate chi(3) | Medium | Ion implantation, CVD | Single electron/ nuclear spins | Strain |
| **Si or GaAs on Rare-earth-doped oxides** | >300nm | | Low | Rare earth incorporation | Electron or nuclear spin qubits | DC Stark Shift, Zeeman Shift |

**IV.B Supporting Technology**

A wide range of supporting technologies will be critical to enabling the variety of QuICs discussed in this document, and their sustained development is crucial. A few examples are highlighted in this section.

Almost all quantum computing and quantum communication approaches—which require the ability to make measurements of a quantum state—use devices that perform best at cryogenic temperatures, where thermal noise can be avoided. For example, in quantum communication systems, optical detectors are an essential component, and cryogenic superconducting nanowire single-photon detectors are currently the state of the art (in terms of efficiency, noise, and timing jitter). In addition, superconducting quantum computing requires operation at ultralow temperatures to maintain qubit integrity. In most of these systems, achieving such low temperatures requires the use of helium. For temperatures below 0.8K, the use of helium-3 is also required. Unfortunately, helium is a strategically important, non-renewable natural resource, and is becoming scarcer.

In the past decade, there has been significant technological progress to use mechanical coolers using recyclable helium gas as a refrigerant to get to sufficiently low temperatures. These new cryo-cooling systems, however, were not developed to meet the needs of the quantum information community. While the cost, size, weight, and price (C-SWaP) are sufficient for research purposes, widespread adoption in commercial applications will be hampered by the high C-SWAP. Convergent research into the development of efficient, long-life, lightweight coolers with low vibration will provide a critical enabling technology for accelerating progress in quantum science and technology.

In addition, the research and development of robust, inexpensive, low-noise, and stable lasers would accelerate both research and commercialization of quantum information science and technology. For optical wavelengths that overlap with existing large markets (e.g. telecommunications), compact lasers already exist. However, for wavelengths that are of interest for atomic and artificial atomic systems (e.g., quantum dots, defects in diamond, SiC), significant effort is spent by the research community to optimize and stabilize custom-built lasers to the level of performance needed to enable quantum applications. Unfortunately, the reliability, stability, and cost of these lasers are not at a level for widespread adoption by researchers or early adopters. There is tremendous opportunity to accelerate progress with multi-disciplinary research into making better-targeted lasers.

A number of additional supporting technologies, including high-speed low-power cryo-compatible classical digital and analog electronics, will also be necessary, and thus warrant development efforts. Similarly, development of non-cryogenic counterparts of currently cryogenic technologies is important. For example, development of single-photon counters such as linear-mode or avalanche-mode photodiodes, could substantially simplify the task of creating scalable repeater technologies [145], [146].

Many of these developments would also benefit classical computation and communication systems, and as such are examples of the *dual-use* paradigm of technology innovation in which quantum-inspired advances assist classical technologies and vice versa.

## V. Conclusions

As the size of quantum systems grows, in terms of number of qubits in the case of quantum computers, or physical size/spatial separation in the case of quantum networks, so do the challenges related to connecting different parts of the system while maintaining quantum entanglement across it. For example, long-range communication networks rely on establishing, distributing and maintaining entanglement across thousands of kilometers. This is challenging due to unavoidable signal losses in the communication channels. At shorter length scales, difficulties associated with connecting hundreds or thousands of qubits point to the importance of modular quantum computing schemes - likely the *only* viable many-qubit approach in the near term. Therefore, QuICs, which will support modular and distributed QIT systems, are emerging as a grand challenge for QIT. Yet, they have received significantly less attention from the funding agencies and from the research community than the quantum hardware systems they are connecting.

It is the position of the community, as represented by participants of the NSF workshop on QuICs, that investment in a national-scale QuIC program is a high priority. Given the diversity of QIT platforms, materials used, applications, and infrastructure required, a convergent research approach and partnership between academia, industry and national laboratories is required for these efforts.

The focus of the envisioned QuICs program should be: (1) a small number of well-supported 'Convergent Development Teams' comprised of specialists from academia, industry, and national laboratories, to address specific QuIC challenges, to create prototype quantum interconnects and application developments; (2) a focused interdisciplinary effort aimed at the development of scalable integrated quantum photonic platforms for QuICs—such an effort should include synthesis of emerging quantum materials, fabrication and packaging of integrated quantum photonic devices, and development of novel ultra-low loss optical fibers; (3) a QuICs Test Bed where researchers would gain access to the equipment and expertise needed to test their own hardware (e.g., qubits, optical squeezing modules, frequency conversion modules, entanglement sources, transducers, sensors, detectors, lasers, etc.), and thus carry out research in a convergent environment. The program will drive the advancement of a quantum-information-science and technology ecosystem that combines research and technology development with commercial and educational elements. This will result in new university degrees (e.g. quantum engineering), creation of student internships in industry, and retraining of current industry employees, thus resulting in an appropriately skilled workforce.

The goals discussed here for accelerating progress in Quantum Interconnects resonate with the goals of two other recent related NSF Accelerator workshops—Quantum Simulators and Quantum Computers.

**Acknowledgement :** The authors acknowledge NSF OIA-1946564 grant "Project Scoping Workshop (PSW) on Quantum Interconnects (QuIC)" that provided financial support for the workshop. The participants are thankful to Ms. Kathleen L. Masse from John A. Paulson School of Engineering at Harvard University for help with organization of the workshop.

**REFERENCES**

[1] National Science and Technology Council, "National Strategic Overview for Quantum Information Science," Sep. 2018.
[2] C. Monroe, M. G. Raymer, and J. Taylor, "The U.S. National Quantum Initiative: From Act to action," *Science*, vol. 364, no. 6439, pp. 440–442, May 2019.
[3] J. Preskill, "Quantum Computing in the NISQ era and beyond," *Quantum*, vol. 2, p. 79, Aug. 2018.
[4] H. J. Kimble, "The quantum internet," *Nature*, vol. 453, no. 7198, pp. 1023–1030, Jun. 2008.
[5] S. Wehner, D. Elkouss, and R. Hanson, "Quantum internet: A vision for the road ahead," *Science*, vol. 362, no. 6412, Oct. 2018.
[6] C. Monroe *et al.*, "Large-scale modular quantum-computer architecture with atomic memory and photonic interconnects," *Phys. Rev. A*, vol. 89, no. 2, p. 022317, Feb. 2014.
[7] D. Gottesman and I. L. Chuang, "Demonstrating the viability of universal quantum computation using teleportation and single-qubit operations," *Nature*, vol. 402, no. 6760, pp. 390–393, Nov. 1999.
[8] J. Eisert, K. Jacobs, P. Papadopoulos, and M. B. Plenio, "Optimal local implementation of nonlocal quantum gates," *Phys. Rev. A*, vol. 62, no. 5, p. 052317, Oct. 2000.
[9] L.-M. Duan, B. B. Blinov, D. L. Moehring, and C. Monroe, "Scalable Trapped Ion Quantum Computation with a Probabilistic Ion-photon Mapping," *Quantum Inf. Comput.*, vol. 4, no. 3, pp. 165–173, May 2004.
[10] L. Jiang, J. M. Taylor, A. S. Sørensen, and M. D. Lukin, "Distributed quantum computation based on small quantum registers," *Phys. Rev. A*, vol. 76, no. 6, p. 062323, Dec. 2007.
[11] J. I. Cirac, P. Zoller, H. J. Kimble, and H. Mabuchi, "Quantum State Transfer and Entanglement Distribution among Distant Nodes in a Quantum Network," *Phys. Rev. Lett.*, vol. 78, no. 16, pp. 3221–3224, Apr. 1997.
[12] S. Ritter *et al.*, "An elementary quantum network of single atoms in optical cavities," *Nature*, vol. 484, no. 7393, pp. 195–200, Apr. 2012.
[13] C. J. Axline *et al.*, "On-demand quantum state transfer and entanglement between remote microwave cavity memories," *Nat. Phys.*, vol. 14, no. 7, pp. 705–710, Apr. 2018.
[14] P. Campagne-Ibarcq *et al.*, "Deterministic Remote Entanglement of Superconducting Circuits through Microwave Two-Photon Transitions," *Phys. Rev. Lett.*, vol. 120, no. 20, p. 200501, May 2018.
[15] P. Kurpiers *et al.*, "Deterministic quantum state transfer and remote entanglement using microwave photons," *Nature*, vol. 558, no. 7709, pp. 264–267, Jun. 2018.
[16] N. Leung *et al.*, "Deterministic bidirectional communication and remote entanglement generation between superconducting qubits," *npj Quantum Information*, vol. 5, no. 1, p. 18, Feb. 2019.
[17] C. Simon and W. T. M. Irvine, "Robust long-distance entanglement and a loophole-free bell test with ions and photons," *Phys. Rev. Lett.*, vol. 91, no. 11, p. 110405, Sep. 2003.
[18] D. L. Moehring *et al.*, "Entanglement of single-atom quantum bits at a distance," *Nature*, vol. 449, no. 7158, pp. 68–71, Sep. 2007.
[19] D. Hucul *et al.*, "Modular entanglement of atomic qubits using photons and phonons," *Nat. Phys.*, vol. 11, no. 1, pp. 37–42, Jan. 2015.
[20] L. J. Stephenson *et al.*, "High-rate, high-fidelity entanglement of qubits across an elementary

quantum network," *arXiv preprint arXiv:1911. 10841*, 2019.
[21] S. Olmschenk, D. N. Matsukevich, P. Maunz, D. Hayes, L.-M. Duan, and C. Monroe, "Quantum teleportation between distant matter qubits," *Science*, vol. 323, no. 5913, pp. 486–489, Jan. 2009.
[22] W. Pfaff *et al.*, "Quantum information. Unconditional quantum teleportation between distant solid-state quantum bits," *Science*, vol. 345, no. 6196, pp. 532–535, Aug. 2014.
[23] J. Hofmann *et al.*, "Heralded Entanglement Between Widely Separated Atoms," *Science*, vol. 337, no. 6090. pp. 72–75, 2012.
[24] H. Bernien *et al.*, "Heralded entanglement between solid-state qubits separated by three metres," *Nature*, vol. 497, no. 7447, pp. 86–90, May 2013.
[25] A. Delteil, Z. Sun, W.-B. Gao, E. Togan, S. Faelt, and A. Imamoğlu, "Generation of heralded entanglement between distant hole spins," *Nat. Phys.*, vol. 12, no. 3, pp. 218–223, Mar. 2016.
[26] A. Reiserer and G. Rempe, "Cavity-based quantum networks with single atoms and optical photons," *Rev. Mod. Phys.*, vol. 87, no. 4, pp. 1379–1418, Dec. 2015.
[27] M. Saffman, "Quantum computing with atomic qubits and Rydberg interactions: progress and challenges," *J. Phys. B At. Mol. Opt. Phys.*, vol. 49, no. 20, p. 202001, Oct. 2016.
[28] C. Monroe and J. Kim, "Scaling the ion trap quantum processor," *Science*, vol. 339, no. 6124, pp. 1164–1169, Mar. 2013.
[29] P. Samutpraphoot *et al.*, "Strong coupling of two individually controlled atoms via a nanophotonic cavity," *arXiv [quant-ph]*, 19-Sep-2019.
[30] M. K. Bhaskar *et al.*, "Experimental demonstration of memory-enhanced quantum communication," *arXiv [quant-ph]*, 03-Sep-2019.
[31] Z. L. Xiang, M. Zhang, L. Jiang, and P. Rabl, "Intracity quantum communication via thermal microwave networks," *Physical Review X*, 2017.
[32] B. Vermersch, P.-O. Guimond, H. Pichler, and P. Zoller, "Quantum State Transfer via Noisy Photonic and Phononic Waveguides," *Phys. Rev. Lett.*, vol. 118, no. 13, p. 133601, Mar. 2017.
[33] K. S. Chou *et al.*, "Deterministic teleportation of a quantum gate between two logical qubits," *Nature*, vol. 561, no. 7723, pp. 368–373, Sep. 2018.
[34] Y. Wan *et al.*, "Quantum gate teleportation between separated qubits in a trapped-ion processor," *Science*, vol. 364, no. 6443, pp. 875–878, May 2019.
[35] S. Pirandola and S. L. Braunstein, "Physics: Unite to build a quantum Internet," *Nature*, vol. 532, no. 7598, pp. 169–171, Apr. 2016.
[36] N. Gisin, G. Ribordy, W. Tittel, and H. Zbinden, "Quantum cryptography," *Reviews of Modern Physics*, vol. 74, no. 1. pp. 145–195, 2002.
[37] H.-K. Lo, X. Ma, and K. Chen, "Decoy State Quantum Key Distribution," *Physical Review Letters*, vol. 94, no. 23. 2005.
[38] H.-K. Lo and N. Lütkenhaus, "Quantum Cryptography: from Theory to Practice," *arXiv [quant-ph]*, 22-Feb-2007.
[39] V. Scarani, H. Bechmann-Pasquinucci, N. J. Cerf, M. Dušek, N. Lütkenhaus, and M. Peev, "The security of practical quantum key distribution," *Reviews of Modern Physics*, vol. 81, no. 3. pp. 1301–1350, 2009.
[40] H.-K. Lo, M. Curty, and B. Qi, "Measurement-Device-Independent Quantum Key Distribution," *Physical Review Letters*, vol. 108, no. 13. 2012.
[41] M. Lucamarini, Z. L. Yuan, J. F. Dynes, and A. J. Shields, "Overcoming the rate–distance limit of quantum key distribution without quantum repeaters," *Nature*, vol. 557, no. 7705, pp. 400–403, May 2018.
[42] S. Pirandola *et al.*, "Advances in Quantum Cryptography," *arXiv [quant-ph]*, 04-Jun-2019.
[43] S. L. Braunstein and S. Pirandola, "Side-channel-free quantum key distribution," *Phys. Rev. Lett.*, vol. 108, no. 13, p. 130502, Mar. 2012.

[44] M. Hillery, V. Bužek, and A. Berthiaume, "Quantum secret sharing," *Phys. Rev. A*, vol. 59, no. 3, pp. 1829–1834, Mar. 1999.
[45] B. P. Williams, J. M. Lukens, N. A. Peters, B. Qi, and W. P. Grice, "Quantum secret sharing with polarization-entangled photon pairs," *Phys. Rev. A*, vol. 99, no. 6, p. 062311, Jun. 2019.
[46] H. Buhrman, R. Cleve, J. Watrous, and R. de Wolf, "Quantum Fingerprinting," *Physical Review Letters*, vol. 87, no. 16. 2001.
[47] J. M. Arrazola and N. Lütkenhaus, "Quantum communication with coherent states and linear optics," *Phys. Rev. A*, vol. 90, no. 4, p. 042335, Oct. 2014.
[48] J.-Y. Guan *et al.*, "Observation of Quantum Fingerprinting Beating the Classical Limit," *Phys. Rev. Lett.*, vol. 116, no. 24, p. 240502, Jun. 2016.
[49] A. Broadbent and C. Schaffner, "Quantum cryptography beyond quantum key distribution," *Des. Codes Cryptogr.*, vol. 78, no. 1, pp. 351–382, Jan. 2016.
[50] A. Broadbent, J. Fitzsimons, and E. Kashefi, "Universal Blind Quantum Computation," in *2009 50th Annual IEEE Symposium on Foundations of Computer Science*, 2009, pp. 517–526.
[51] J. F. Fitzsimons, "Private quantum computation: an introduction to blind quantum computing and related protocols," *npj Quantum Information*, vol. 3, no. 1, p. 23, Jun. 2017.
[52] W. T. Buttler *et al.*, "Practical Free-Space Quantum Key Distribution over 1 km," *Phys. Rev. Lett.*, vol. 81, no. 15, pp. 3283–3286, Oct. 1998.
[53] X.-S. Ma *et al.*, "Quantum teleportation over 143 kilometres using active feed-forward," *Nature*, vol. 489, no. 7415, pp. 269–273, Sep. 2012.
[54] S.-K. Liao *et al.*, "Long-distance free-space quantum key distribution in daylight towards inter-satellite communication," *Nature Photonics*, vol. 11, no. 8. pp. 509–513, 2017.
[55] F. Steinlechner *et al.*, "Distribution of high-dimensional entanglement via an intra-city free-space link," *Nat. Commun.*, vol. 8, p. 15971, Jul. 2017.
[56] R. Valivarthi *et al.*, "Quantum teleportation across a metropolitan fibre network," *Nature Photonics*, vol. 10, no. 10. pp. 676–680, 2016.
[57] Q.-C. Sun *et al.*, "Quantum teleportation with independent sources and prior entanglement distribution over a network," *Nat. Photonics*, vol. 10, no. 10, pp. 671–675, Oct. 2016.
[58] C. J. Pugh *et al.*, "Airborne demonstration of a quantum key distribution receiver payload," *Quantum Science and Technology*, vol. 2, no. 2. p. 024009, 2017.
[59] H.-Y. Liu *et al.*, "Drone-based all-weather entanglement distribution," *arXiv [quant-ph]*, 23-May-2019.
[60] J.-G. Ren *et al.*, "Ground-to-satellite quantum teleportation," *Nature*, vol. 549, no. 7670, pp. 70–73, Sep. 2017.
[61] S.-K. Liao *et al.*, "Satellite-Relayed Intercontinental Quantum Network," *Phys. Rev. Lett.*, vol. 120, no. 3, p. 030501, Jan. 2018.
[62] H.-J. Briegel, W. Dür, J. I. Cirac, and P. Zoller, "Quantum Repeaters: The Role of Imperfect Local Operations in Quantum Communication," *Phys. Rev. Lett.*, vol. 81, no. 26, pp. 5932–5935, Dec. 1998.
[63] L. M. Duan, M. D. Lukin, J. I. Cirac, and P. Zoller, "Long-distance quantum communication with atomic ensembles and linear optics," *Nature*, vol. 414, no. 6862, pp. 413–418, Nov. 2001.
[64] L. Jiang, J. M. Taylor, K. Nemoto, W. J. Munro, R. Van Meter, and M. D. Lukin, "Quantum repeater with encoding," *Phys. Rev. A*, vol. 79, no. 3, p. 032325, Mar. 2009.
[65] W. J. Munro, K. A. Harrison, A. M. Stephens, S. J. Devitt, and K. Nemoto, "From quantum multiplexing to high-performance quantum networking," *Nat. Photonics*, vol. 4, no. 11, pp. 792–796, Nov. 2010.
[66] W. J. Munro, A. M. Stephens, S. J. Devitt, K. A. Harrison, and K. Nemoto, "Quantum communication without the necessity of quantum memories," *Nat. Photonics*, vol. 6, no. 11, pp.

777–781, Nov. 2012.
[67] S. Muralidharan, J. Kim, N. Lütkenhaus, M. D. Lukin, and L. Jiang, "Ultrafast and fault-tolerant quantum communication across long distances," *Phys. Rev. Lett.*, vol. 112, no. 25, p. 250501, Jun. 2014.
[68] M. Takeoka, S. Guha, and M. M. Wilde, "Fundamental rate-loss tradeoff for optical quantum key distribution," *Nat. Commun.*, vol. 5, p. 5235, Oct. 2014.
[69] S. Pirandola, R. Laurenza, C. Ottaviani, and L. Banchi, "Fundamental limits of repeaterless quantum communications," *Nat. Commun.*, vol. 8, p. 15043, Apr. 2017.
[70] S. G. Carter *et al.*, "Quantum control of a spin qubit coupled to a photonic crystal cavity," *Nat. Photonics*, vol. 7, no. 4, pp. 329–334, Apr. 2013.
[71] C. T. Nguyen *et al.*, "Quantum Network Nodes Based on Diamond Qubits with an Efficient Nanophotonic Interface," *Phys. Rev. Lett.*, vol. 123, no. 18, p. 183602, Nov. 2019.
[72] F. Kaneda, F. Xu, J. Chapman, and P. G. Kwiat, "Quantum-memory-assisted multi-photon generation for efficient quantum information processing," *Optica, OPTICA*, vol. 4, no. 9, pp. 1034–1037, Sep. 2017.
[73] H. Jeong *et al.*, "Generation of hybrid entanglement of light," *Nat. Photonics*, vol. 8, no. 7, pp. 564–569, Jul. 2014.
[74] J. T. Barreiro, N. K. Langford, N. A. Peters, and P. G. Kwiat, "Generation of hyperentangled photon pairs," *Phys. Rev. Lett.*, vol. 95, no. 26, p. 260501, Dec. 2005.
[75] P. Imany *et al.*, "High-dimensional optical quantum logic in large operational spaces," *npj Quantum Information*, vol. 5, no. 1, p. 59, Jul. 2019.
[76] C. Reimer *et al.*, "High-dimensional one-way quantum processing implemented on d-level cluster states," *Nat. Phys.*, vol. 15, no. 2, pp. 148–153, Dec. 2018.
[77] V. B. Braginsky, Y. I. Vorontsov, and K. S. Thorne, "Quantum nondemolition measurements," *Science*, vol. 209, no. 4456, pp. 547–557, Aug. 1980.
[78] G. Santarelli *et al.*, "Quantum Projection Noise in an Atomic Fountain: A High Stability Cesium Frequency Standard," *Phys. Rev. Lett.*, vol. 82, no. 23, pp. 4619–4622, Jun. 1999.
[79] W. Wasilewski, K. Jensen, H. Krauter, J. J. Renema, M. V. Balabas, and E. S. Polzik, "Quantum noise limited and entanglement-assisted magnetometry," *Phys. Rev. Lett.*, vol. 104, no. 13, p. 133601, Apr. 2010.
[80] P. H. Kim, B. D. Hauer, C. Doolin, F. Souris, and J. P. Davis, "Approaching the standard quantum limit of mechanical torque sensing," *Nat. Commun.*, vol. 7, p. 13165, Oct. 2016.
[81] M. Tse *et al.*, "Quantum-Enhanced Advanced LIGO Detectors in the Era of Gravitational-Wave Astronomy," *Phys. Rev. Lett.*, vol. 123, no. 23, p. 231107, Dec. 2019.
[82] J. R. Maze *et al.*, "Nanoscale magnetic sensing with an individual electronic spin in diamond," *Nature*, vol. 455, no. 7213, pp. 644–647, Oct. 2008.
[83] Z. Zhang, S. Mouradian, F. N. C. Wong, and J. H. Shapiro, "Entanglement-enhanced sensing in a lossy and noisy environment," *Phys. Rev. Lett.*, vol. 114, no. 11, p. 110506, Mar. 2015.
[84] J. Appel, P. J. Windpassinger, D. Oblak, U. B. Hoff, N. Kjaergaard, and E. S. Polzik, "Mesoscopic atomic entanglement for precision measurements beyond the standard quantum limit," *Proc. Natl. Acad. Sci. U. S. A.*, vol. 106, no. 27, pp. 10960–10965, Jul. 2009.
[85] J. D. Teufel, T. Donner, M. A. Castellanos-Beltran, J. W. Harlow, and K. W. Lehnert, "Nanomechanical motion measured with an imprecision below that at the standard quantum limit," *Nat. Nanotechnol.*, vol. 4, no. 12, pp. 820–823, Dec. 2009.
[86] A. Mainwaring, D. Culler, and J. Polastre, "Wireless sensor networks for habitat monitoring," *Proceedings of the 1st*, 2002.
[87] R. C. Hansen, *Phased Array Antennas*. John Wiley & Sons, 2009.
[88] E. T. Khabiboulline, J. Borregaard, K. De Greve, and M. D. Lukin, "Optical Interferometry with

Quantum Networks," *Phys. Rev. Lett.*, vol. 123, no. 7, p. 070504, Aug. 2019.

[89] P. Kómár et al., "A quantum network of clocks," *Nat. Phys.*, vol. 10, no. 8, pp. 582–587, Aug. 2014.

[90] W. Ge, K. Jacobs, Z. Eldredge, A. V. Gorshkov, and M. Foss-Feig, "Distributed Quantum Metrology with Linear Networks and Separable Inputs," *Phys. Rev. Lett.*, vol. 121, no. 4, p. 043604, Jul. 2018.

[91] T. J. Proctor, P. A. Knott, and J. A. Dunningham, "Multiparameter Estimation in Networked Quantum Sensors," *Phys. Rev. Lett.*, vol. 120, no. 8, p. 080501, Feb. 2018.

[92] Q. Zhuang, Z. Zhang, and J. H. Shapiro, "Distributed quantum sensing using continuous-variable multipartite entanglement," *Phys. Rev. A*, vol. 97, no. 3, p. 032329, Mar. 2018.

[93] Y. Xia, W. Li, W. Clark, D. Hart, Q. Zhuang, and Z. Zhang, "Entangled Radiofrequency-Photonic Sensor Network," *arXiv [quant-ph]*, 19-Oct-2019.

[94] D. Gottesman, T. Jennewein, and S. Croke, "Longer-baseline telescopes using quantum repeaters," *Phys. Rev. Lett.*, vol. 109, no. 7, p. 070503, Aug. 2012.

[95] C. Wang et al., "Integrated lithium niobate electro-optic modulators operating at CMOS-compatible voltages," *Nature*, vol. 562, no. 7725, pp. 101–104, Oct. 2018.

[96] Q. Zhuang and Z. Zhang, "Physical-Layer Supervised Learning Assisted by an Entangled Sensor Network," *Phys. Rev. X*, vol. 9, no. 4, p. 041023, Oct. 2019.

[97] D. DeMille, "Quantum computation with trapped polar molecules," *Phys. Rev. Lett.*, vol. 88, no. 6, p. 067901, Feb. 2002.

[98] P. Lodahl, S. Mahmoodian, and S. Stobbe, "Interfacing single photons and single quantum dots with photonic nanostructures," *Rev. Mod. Phys.*, vol. 87, no. 2, pp. 347–400, May 2015.

[99] A. Sipahigil et al., "An integrated diamond nanophotonics platform for quantum-optical networks," *Science*, vol. 354, no. 6314, pp. 847–850, Nov. 2016.

[100] D. M. Lukin et al., "4H-silicon-carbide-on-insulator for integrated quantum and nonlinear photonics," *Nat. Photonics*, Dec. 2019.

[101] C. Simon et al., "Quantum memories," *Eur. Phys. J. D*, vol. 58, no. 1, pp. 1–22, May 2010.

[102] A. M. Dibos, M. Raha, C. M. Phenicie, and J. D. Thompson, "Atomic Source of Single Photons in the Telecom Band," *Phys. Rev. Lett.*, vol. 120, no. 24, p. 243601, Jun. 2018.

[103] M. Businger et al., "Optical spin-wave storage in a solid-state hybridized electron-nuclear spin ensemble," *arXiv [quant-ph]*, 26-Jul-2019.

[104] T. Zhong et al., "Nanophotonic rare-earth quantum memory with optically controlled retrieval," *Science*, vol. 357, no. 6358, pp. 1392–1395, Sep. 2017.

[105] M. P. Hedges, J. J. Longdell, Y. Li, and M. J. Sellars, "Efficient quantum memory for light," *Nature*, vol. 465, no. 7301, pp. 1052–1056, Jun. 2010.

[106] T. Zhong et al., "Optically Addressing Single Rare-Earth Ions in a Nanophotonic Cavity," *Phys. Rev. Lett.*, vol. 121, no. 18, p. 183603, Nov. 2018.

[107] M. H. Devoret and R. J. Schoelkopf, "Superconducting circuits for quantum information: an outlook," *Science*, vol. 339, no. 6124, pp. 1169–1174, Mar. 2013.

[108] I. Aharonovich, D. Englund, and M. Toth, "Solid-state single-photon emitters," *Nat. Photonics*, vol. 10, no. 10, pp. 631–641, Oct. 2016.

[109] K. Heshami et al., "Quantum memories: emerging applications and recent advances," *J. Mod. Opt.*, vol. 63, no. 20, pp. 2005–2028, Nov. 2016.

[110] M. Pechal and A. H. Safavi-Naeini, "Millimeter-wave interconnects for microwave-frequency quantum machines," *Phys. Rev. A*, vol. 96, no. 4, p. 042305, Oct. 2017.

[111] P. Kurpiers, T. Walter, P. Magnard, Y. Salathe, and A. Wallraff, "Characterizing the attenuation of coaxial and rectangular microwave-frequency waveguides at cryogenic temperatures," *EPJ Quantum Technol*, vol. 4, no. 1, p. 8, May 2017.

[112] A. H. Safavi-Naeini, D. Van Thourhout, R. Baets, and R. Van Laer, "Controlling phonons and photons at the wavelength scale: integrated photonics meets integrated phononics," *Optica, OPTICA*,

vol. 6, no. 2, pp. 213–232, Feb. 2019.
[113] J. Huang and P. Kumar, "Observation of quantum frequency conversion," *Phys. Rev. Lett.*, vol. 68, no. 14, pp. 2153–2156, Apr. 1992.
[114] A. P. VanDevender and P. G. Kwiat, "Quantum transduction via frequency upconversion (Invited)," *J. Opt. Soc. Am. B, JOSAB*, vol. 24, no. 2, pp. 295–299, Feb. 2007.
[115] M. Bock *et al.*, "High-fidelity entanglement between a trapped ion and a telecom photon via quantum frequency conversion," *Nat. Commun.*, vol. 9, no. 1, p. 1998, May 2018.
[116] A. Singh *et al.*, "Quantum frequency conversion of a quantum dot single-photon source on a nanophotonic chip," *Optica*, vol. 6, no. 5. p. 563, 2019.
[117] L. J. Wright, M. Karpiński, C. Söller, and B. J. Smith, "Spectral Shearing of Quantum Light Pulses by Electro-Optic Phase Modulation," *Phys. Rev. Lett.*, vol. 118, no. 2, p. 023601, Jan. 2017.
[118] M. G. Raymer and K. Srinivasan, "Manipulating the color and shape of single photons," *Phys. Today*, vol. 65, no. 11, p. 32, 2012.
[119] R. Salem, M. A. Foster, A. C. Turner, D. F. Geraghty, M. Lipson, and A. L. Gaeta, "Optical time lens based on four-wave mixing on a silicon chip," *Opt. Lett.*, vol. 33, no. 10, pp. 1047–1049, May 2008.
[120] B. Brecht, D. V. Reddy, C. Silberhorn, and M. G. Raymer, "Photon Temporal Modes: A Complete Framework for Quantum Information Science," *Phys. Rev. X*, vol. 5, no. 4, p. 041017, Oct. 2015.
[121] K. Banaszek, M. Jachura, and W. Wasilewski, "Utilizing time-bandwidth space for efficient deep-space communication," in *International Conference on Space Optics — ICSO 2018*, 2019, vol. 11180, p. 111805X.
[122] F. Arute *et al.*, "Quantum supremacy using a programmable superconducting processor," *Nature*, vol. 574, no. 7779, pp. 505–510, Oct. 2019.
[123] Y.-I. Sohn *et al.*, "Controlling the coherence of a diamond spin qubit through its strain environment," *Nat. Commun.*, vol. 9, no. 1, p. 2012, May 2018.
[124] M.-A. Lemonde *et al.*, "Phonon Networks with Silicon-Vacancy Centers in Diamond Waveguides," *Phys. Rev. Lett.*, vol. 120, no. 21, p. 213603, May 2018.
[125] M. Atatüre, D. Englund, N. Vamivakas, S.-Y. Lee, and J. Wrachtrup, "Material platforms for spin-based photonic quantum technologies," *Nature Reviews Materials*, vol. 3, no. 5. pp. 38–51, 2018.
[126] S. L. Mouradian *et al.*, "Scalable Integration of Long-Lived Quantum Memories into a Photonic Circuit," *Phys. Rev. X*, vol. 5, no. 3, p. 031009, Jul. 2015.
[127] D. S. Hum and M. M. Fejer, "Quasi-phasematching," *C. R. Phys.*, vol. 8, no. 2, pp. 180–198, Mar. 2007.
[128] C. Wang *et al.*, "Ultrahigh-efficiency wavelength conversion in nanophotonic periodically poled lithium niobate waveguides," *Optica, OPTICA*, vol. 5, no. 11, pp. 1438–1441, Nov. 2018.
[129] J. Lu *et al.*, "Periodically poled thin-film lithium niobate microring resonators with a second-harmonic generation efficiency of 250,000%/W," *Optica, OPTICA*, vol. 6, no. 12, pp. 1455–1460, Dec. 2019.
[130] X. Guo, C.-L. Zou, H. Jung, and H. X. Tang, "On-Chip Strong Coupling and Efficient Frequency Conversion between Telecom and Visible Optical Modes," *Phys. Rev. Lett.*, vol. 117, no. 12, p. 123902, Sep. 2016.
[131] M. Zhang, C. Wang, R. Cheng, A. Shams-Ansari, and M. Lončar, "Monolithic ultra-high-Q lithium niobate microring resonator," *Optica, OPTICA*, vol. 4, no. 12, pp. 1536–1537, Dec. 2017.
[132] B. Desiatov, A. Shams-Ansari, M. Zhang, C. Wang, and M. Lončar, "Ultra-low-loss integrated visible photonics using thin-film lithium niobate," *Optica, OPTICA*, vol. 6, no. 3, pp. 380–384, Mar. 2019.

[133] Q. Li *et al.*, "Tunable Quantum Beat of Single Photons Enabled by Nonlinear Nanophotonics," *Phys. Rev. Applied*, vol. 12, no. 5, p. 054054, Nov. 2019.
[134] D. J. Wilson *et al.*, "Integrated gallium phosphide nonlinear photonics," *Nature Photonics*. 2019.
[135] B. J. M. Hausmann, I. Bulu, V. Venkataraman, P. Deotare, and M. Lončar, "Diamond nonlinear photonics," *Nat. Photonics*, vol. 8, no. 5, pp. 369–374, May 2014.
[136] X. Lu *et al.*, "Chip-integrated visible-telecom photon pair sources for quantum communication," *Nat. Phys.*, vol. 15, 2019.
[137] J. S. Pelc *et al.*, "Long-wavelength-pumped upconversion single-photon detector at 1550 nm: performance and noise analysis," *Opt. Express*, vol. 19, no. 22, pp. 21445–21456, Oct. 2011.
[138] J. S. Pelc, C. R. Phillips, D. Chang, C. Langrock, and M. M. Fejer, "Efficiency pedestal in quasi-phase-matching devices with random duty-cycle errors," *Opt. Lett.*, vol. 36, no. 6, pp. 864–866, Mar. 2011.
[139] L. Fan *et al.*, "Superconducting cavity electro-optics: A platform for coherent photon conversion between superconducting and photonic circuits," *Sci Adv*, vol. 4, no. 8, p. eaar4994, Aug. 2018.
[140] L. Fan *et al.*, "Cavity electro-optic circuit for microwave-to-optical frequency conversion," in *Nonlinear Optics (NLO)*, Waikoloa Beach, Hawaii United States, 2019, p. NF2A.2.
[141] A. P. Higginbotham *et al.*, "Harnessing electro-optic correlations in an efficient mechanical converter," *Nat. Phys.*, vol. 14, no. 10, pp. 1038–1042, Oct. 2018.
[142] L. A. Williamson, Y.-H. Chen, and J. J. Longdell, "Magneto-optic modulator with unit quantum efficiency," *Phys. Rev. Lett.*, vol. 113, no. 20, p. 203601, Nov. 2014.
[143] J. G. Bartholomew *et al.*, "On-chip coherent microwave-to-optical transduction mediated by ytterbium in YVO $\_4$," *arXiv preprint arXiv:1912. 03671*, 2019.
[144] T. Komljenovic *et al.*, "Heterogeneous Silicon Photonic Integrated Circuits," *J. Lightwave Technol., JLT*, vol. 34, no. 1, pp. 20–35, Jan. 2016.
[145] K. Azuma, K. Tamaki, and H.-K. Lo, "All-photonic quantum repeaters," *Nat. Commun.*, vol. 6, p. 6787, Apr. 2015.
[146] M. Pant, H. Krovi, D. Englund, and S. Guha, "Rate-distance tradeoff and resource costs for all-optical quantum repeaters," *Phys. Rev. A*, vol. 95, no. 1, p. 012304, Jan. 2017.